%% file: paper.tex
\documentclass{llncs}
\pdfoutput=1
\usepackage{url}
\usepackage{hyperref}
 \hypersetup{
 	colorlinks,
 	citecolor=red,
 	filecolor=blue,
 	linkcolor=darkgray,
 	urlcolor=red
 } 

\usepackage{outlines}

\usepackage[%
    font={small,sf},
    labelfont=bf,
    format=hang,    
    format=plain,
    margin=0pt,
    width=\textwidth,
]{caption}
\usepackage{subcaption}
\usepackage[list=true]{subcaption}
\usepackage{paralist}
\usepackage{color}
\usepackage{xcolor}
\usepackage{wrapfig } 
\usepackage{booktabs}
\usepackage{multido} 
\usepackage{multirow}
\usepackage{bigstrut}
\usepackage{longtable}
\usepackage[inline, shortlabels]{enumitem} 
\usepackage{rotating}
\usepackage{lipsum} 
\usepackage{graphicx}
\usepackage{subcaption}
\usepackage[colorinlistoftodos]{todonotes}
\usepackage{xspace} 
\usepackage{framed} 
\usepackage{mdframed}
\usepackage[normalem]{ulem}

\usepackage{cooltooltips} 

\usepackage[hybrid,fencedCode]{markdown} 

\newcommand{\todoALL}[1]{\todo[linecolor=red!70!white,color=red!40]{@All: #1}}

\newcommand{\todoVK} [1]{\todo[linecolor=orange!70!white,color=orange!40]{#1}}

\definecolor{jonas}{RGB}{128, 1, 0}

\newcommand{\openmp}{Open{MP}\xspace}

\usepackage{abstract}

\usepackage{listings}

\usepackage{tikz}                        
\usetikzlibrary{arrows,automata,calc,positioning,shapes,shadows.blur,decorations.pathreplacing,decorations.markings}

\newcommand{\comments}[1]{}
\begin{document}

\title{Toward A Standard Interface for \\User-Defined Scheduling in \openmp}

\pagestyle{headings}
\pagenumbering{arabic}

\titlerunning{UDS in OpenMP}  
%
\author{Vivek~Kale\inst{1}\and Christian~Iwainsky\inst{2}\and Michael~Klemm\inst{3}\and Jonas~H.~M\"uller Kornd\"orfer\inst{4}\and Florina~M.~Ciorba\inst{4}}

\authorrunning{X et al.} 
%
\tocauthor{Vivek Kale, Christian Iwainsky, Michael Klemm, \\Jonas H. M\"uller-Kornd\"orfer, and Florina M. Ciorba}
\institute{ 
Brookhaven National Laboratory, Upton, USA\\
\and
Technische Universit\"at Darmstadt, Darmstadt, Germany\\
\and
Intel Deutschland GmbH, Feldkirchen, Germany\\
\and
University of Basel, Basel, Switzerland
}
\setcounter{footnote}{0}
\maketitle              
\setcounter{footnote}{0}
\begin{abstract}
Parallel loops are an important part of \openmp programs. Efficient
scheduling of parallel loops can improve performance of the
programs. The current \openmp specification only offers three options
for loop scheduling, which are insufficient in certain
instances. Given the large number of other possible scheduling
strategies, standardizing each of them is infeasible. A more viable
approach is to extend the \openmp standard to allow a user to define
loop scheduling strategies within her application. The approach will enable
standard-compliant application-specific scheduling.
This work analyzes the principal components required by user-defined
scheduling and proposes two competing interfaces as candidates for the
\openmp standard. We conceptually compare the two proposed interfaces
with respect to the three host languages of \openmp, i.e., C, C++, and
Fortran. These interfaces serve the \openmp community as a basis for
discussion and prototype implementation supporting user-defined scheduling 
in an \openmp library.

\end{abstract}

\section{Introduction}
\label{sec:intro}
\input{intro2.0}
\vspace{-.25cm}
\section{Scheduling Background and State of the Art}
\label{sec:bkgrnd-sota}
\input{bkgrnd-sota}
\vspace{-.25cm}
\section{Support for User-defined Scheduling Strategies}
\label{sec:rq-analysis}
\input{rq-analysis}

\vspace{-.25cm}
\section{An Interface for User-defined Loop Scheduling}
\vspace*{-.25cm}
%
\input{uds-proposal}
\label{sec:uds-interface}
\subsection{Discussion}
\label{sec:discussion}
\input{discussion}
%

\input{relatedWork}
\vspace{-.25cm}
\section{Conclusion}
\label{sec:conclusion}
\input{conclusion}

\vspace{-.25cm}
\input{ack.tex}
\vspace{-.25cm}
\bibliographystyle{splncs04}
\bibliography{iwomp19-bibliography}

\end{document}

%% file: intro2.0.tex

\openmp\,\cite{OpenMP} is the industry and academic standard for
parallel programming on shared memory platforms.
Loop-level parallelism is a very important part of many
\openmp applications that frequently contain
\emph{computationally-intensive} and large \emph{data
parallel loops}. 
Such \openmp applications are typically executed on high performance
computing (HPC) platforms which are increasingly complex, large,
heterogeneous, and exhibit massive and diverse parallelism. 
The performance of applications executing on HPC platforms can be
degraded due to various \emph{overheads}, such as synchronization,
management of parallelism, communication, and load
imbalance\,\cite{Banicescu:1996}.
Indeed, these overheads cannot be ignored by any effort to improve the
performance of applications, such as the loop scheduling
schemes\,\cite{BAL:1998}.
The scheduling of those large and complex \openmp loops can be a
critical factor for the efficient use of those HPC platforms.

The optimal scheduling of parallel applications on parallel
computing platforms is \mbox{NP-hard}\,\cite{NP-Hard:1990}.
No single loop scheduling technique can address all sources
of \emph{load imbalance} to effectively optimize the performance of
all \emph{parallel applications} executing on all types
of \emph{computing platforms}. Indeed, the characteristics of the loop iterations \emph{compounded}
with the characteristics of the underlying computing systems
determine, typically during execution, whether a certain scheduling
scheme outperforms another. 
The performance of parallel applications is impacted by system-induced variability (e.g., operating system
noise, power capping) and results in additional irregularity that has often been
neglected in loop scheduling research, particularly in the context of
OpenMP scheduling\,\cite{DLS+API:2003,Automatic-OMP-LS:2012}. 
Efficient loop scheduling can mitigate those variabilities, if a suitable
schedule is available. However, choices for loop scheduling strategies in \openmp are limited
today to static, guided, or dynamic.
These three scheduling strategies have been shown in previous
work\,\cite{Ciorba:2018,Kasielke:2019} not to offer the best
performance possible.
Moreover, fault-tolerant and energy-oriented \openmp loop scheduling
strategies require domain-specific knowledge to maintain correctness
and energy-efficiency at large-scale, respectively\,\cite{openmplowpower,openmpft}, which is currently not exploited by the three standard \openmp scheduling strategies.

More and novel loop scheduling strategies are needed in \openmp given
complexity of emerging applications and of supercomputer architectures. 
This is evident by the efforts of compiler developers, open-source and
commercial alike, to support additional scheduling schemes.
The efforts can be observed in LLVM\,\cite{LLVMOpenMP} with the trapezoid
self-scheduling\,\cite{TSS:1993} strategy, or in the Intel compiler with a
static stealing scheme\,\cite{FSC:1985}.
However, given the great body of work on loop scheduling, in general,
standardizing all possible scheduling strategies in \openmp is infeasible.
Therefore, given the many different compilers supporting \openmp, a
standardized way of supporting additional scheduling strategies is mandatory for
portability and use in today's frequently changing HPC landscape.

A more viable approach is to extend the \openmp
standard to allow for \emph{user-defined loop scheduling} (UDS).
Doing so will enable application-specific scheduling as well as
a standard-compliant means to customize current loop schedulers.
To this end, this work analyzes the principal operations of a loop
scheduling scheme using a `todo list' as a representation of the loop
iteration space. Based on this modeling we identify four mandatory
operations (init, enqueue, dequeue, and finalize).
To support all currently available scheduling strategies, additional
information may be necessary, which can be obtained through two
measurement operations around the loop body. 
Using these principal components, we propose two complementary UDS
specification interfaces for \openmp, following the distinct styles of
C, Fortran, and C++. 
One proposal supports a more modern programming style, such as that
used in C++14 and later. 
The other proposal takes a classic approach, is suitable for C, Fortran and C++ programs
and helps many types of applications to run on various architectures.  
The aim is that these proposals serve
the \openmp community and compiler developers as a basis for
discussion and prototype implementation of UDS. 

The core contributions of this work are: 
\begin{inparaenum}
\item[(1)] an analysis of existing scheduling strategies and specifications of a minimal function set that is capable of implementing them and
\item[(2)] an actual language-specific proposal of how to implement existing and future user-defined scheduling strategies.
\end{inparaenum}

The remainder of this paper is structured as follows. First, we
provide the background and state of the art in recent loop scheduling
strategies in Section~\ref{sec:bkgrnd-sota}. 
We then introduce our proposal for an interface in \openmp to
facilitate user-defined scheduling and its design rationale in
Section~\ref{sec:rq-analysis}. We present in
Section~\ref{sec:uds-interface} the two alternative proposals for the
specification of user-defined loop scheduling for \openmp. Finally, we
summarize our experience in Section~\ref{sec:conclusion}.
\footnote{The final publication is available at \url{https://www.springerlink.com}.}

%% file: bkgrnd-sota.tex

Scheduling, as broadly understood, refers to the orchestration of units
of work onto units of execution, in space and time.  It typically
consists of three steps: partitioning, assignment, and load
balancing. 
A computational application is \emph{partitioned} into units of work to expose the software parallelism. 
This parallelism is expressed by \emph{assigning} the units of work
(e.g., problem sub-domains) to units of processing (e.g., processes,
threads, tasks). 
The parallel units of processing are subsequently \emph{assigned} to
units of execution (e.g., nodes, processors, cores) to exploit the
available hardware parallelism. 
\emph{Load balancing} refers to evenly assigning the units of work to
units of processing (software load balancing) or to evenly assigning
the units of processing to units of execution (hardware load
balancing). In load balancing, the transfer policy determines \emph{whether} a
unit of work should be transferred, while the location policy
determines \emph{where} it should be transferred. 
Based on the location policy, load balancing approaches can
be \emph{sender-initiated} (also called work
sharing), \emph{receiver-initiated} (also referred to as
self-scheduling or work stealing),
or \emph{symmetrically-initiated}~\cite{Krueger:1994}.

{Load imbalance} is the major performance degradation overhead in
com\-pu\-ta\-tion\-ally-intensive
applications\,\cite{IESP-2.0,FRAC:1996}. 
It can result from the uneven assignment of units of computation to
units of processing (e.g., threads) or the uneven assignment of units
of processing to units of execution. At light and moderate load
imbalance, sender-initiated and symmetrically-initiated algorithms outperform
receiver-initiated algorithms. Conversely, at high loads, they perform poorly, possibly causing
system instability and are outperformed by receiver-initiated
algorithms~\cite{Krueger:1994}. A \emph{load balanced execution}
refers to the case when all units of
execution complete their assigned work \emph{at the same time}. 

In this work, we concentrate on the scheduling and (software) load
balancing of parallel \openmp loops. In this context, we consider
computational problems that contain parallel loops expressed
using \openmp \emph{worksharing} constructs.
The iterations of these loops are scheduled and load balanced,
respectively, to achieve a load balanced execution.

It is important to note that many scientific, engineering, and
industrial applications that use \openmp contain worksharing loops. 
Therefore, scheduling of worksharing loops in \openmp is not
overshadowed by the recent advances and developments in \openmp
tasking. Worksharing loops and tasking represent two complementary
parallel programming approaches that intersect when each iteration of
a worksharing loop creates an \openmp task to execute the loop body. 

%
%
The term \emph{loop scheduling strategy} denotes the technique (or
policy) for assigning the loop iterations to threads in a team.  
A \emph{loop scheduler} refers to the implementation of a particular
loop scheduling strategy, while \emph{loop schedule} represents the
resulting assignment of loop iterations to threads in a team based on
the particular scheduling strategy and its corresponding scheduler. 
In this work, the acronym UDS denotes \emph{user-defined loop
scheduling}. However, unless otherwise noted, the term UDS is also interchangeably used to denote either scheduling, scheduler, or schedule.


There exists a great body of work on loop scheduling and a taxonomy of
loop scheduling strategies can be found in recent
literature~\cite{Ciorba:2018}. Loop scheduling strategies can broadly
be classified into \emph{static} and \emph{dynamic}. 
The dynamic strategies can further be classified
into \emph{non-adaptive} and \emph{adaptive}. 
The static scheduling strategies take the partitioning, assignment,
and load balancing decisions \emph{before} the loop executes, while dynamic
scheduling strategies take most of or all these decisions during execution. 
Moreover, the dynamic adaptive scheduling strategies \emph{adapt}
these decisions as the loop executes based on the application,
execution, and system states, to deliver a highly balanced execution. 


The \openmp specification\,\cite{OpenMP} offers three scheduling
options for worksharing loops: \texttt{static}, \texttt{dynamic},
and \texttt{guided}.  Each can be directly selected as arguments to
the OpenMP \texttt{schedule()} clause of a \texttt{for} directive.
The first option falls into the \emph{static} scheduling category,
while the other two options belong to the \emph{dynamic non-adaptive}
scheduling category with \emph{receiver-initiated} load balancing
location policy. The loop scheduling strategies can also automatically be selected
by the \openmp runtime system via the \texttt{auto} argument
to \texttt{schedule()} or their selection can be deferred to execution
time via the \texttt{runtime} argument to \texttt{schedule()}.

%
%
The use of \texttt{schedule(static,chunk)}
employs \emph{straightforward parallelization} or \emph{static block
scheduling}\,\cite{LDS:1993} (STATIC) wherein $N$ loop iterations are
divided into $P$ chunks of size $\lceil N/ P \rceil$; $P$ being the
number of units of processing (e.g., threads). Each \texttt{chunk} of consecutive
iterations is assigned to a thread, in a round-robin fashion.
This is only suitable for \emph{uniformly distributed} loop iterations
and in the \emph{absence} of load imbalance. The use
of \texttt{schedule(static,1)} implements \emph{static cyclic
scheduling}\,\cite{LDS:1993} wherein single iterations are statically
assigned consecutively to different threads in a cyclic fashion,
i.e., iteration $i$ is assigned to thread $i$ \texttt{mod} $P$. 
For certain non-uniformly distributed parallel loop iterations, cyclic scheduling
produces a more balanced schedule than block scheduling. 
Both versions achieve high locality with virtually no scheduling
overhead, at the expense of poor load balancing if applied to loops
with irregular loop iterations or in systems with high variability.

The dynamic version of \texttt{schedule(static,}\texttt{chunk)} that
employs \emph{dynamic block scheduling} is \texttt{schedule(dynamic,chunk)}.
It differs in that the assignment of chunks to threads is
performed during execution. The dynamic counterpart to \texttt{schedule(static,1)}
is \texttt{schedule(dynamic,1)} which
employs \emph{pure} \emph{\mbox{self-scheduling}} (PSS or simply SS), the easiest 
and most straightforward dynamic loop \mbox{self-scheduling}
algorithm\,\cite{PSS:1986}. Whenever a thread is idle, it retrieves
an iteration from a central work queue (receiver-initiated load
balancing). SS~achieves good load balancing yet may
cause excessive scheduling overhead.
%
The scheduling option \texttt{schedule(guided)} implements \emph{guided} \emph{\mbox{self-scheduling}}(GSS)\,\cite{GSS:1987}, one of the early self-scheduling-based
techniques that trades off load imbalance and scheduling overhead.

Further noteworthy dynamic non-adaptive loop scheduling techniques
are \emph{trapezoid self-scheduling}
(TSS)\,\cite{TSS:1993}, \emph{factoring2} (FAC2)\,\cite{FAC:1992},
and \emph{weighted factoring2} (WF2)\,\cite{WF:1996}.
TSS, FAC2, and WF2 do not require additional information about loop
characteristics and the allocated chunk sizes using these techniques
decrease during the course of the execution from one work request to
another. It is important to note that the FAC2 and WF2 evolved from
the probabilistic analyses that conceived FAC\,\cite{FAC:1992} and
WF\,\cite{WF:1996}, respectively, while TSS is a deterministic
self-scheduling method. Moreover, WF2 can employ workload balancing 
information specified by the user, such as the capabilities of a
heterogeneous hardware 
configuration. 

TSS, FAC2, WF2, and RAND (random self-scheduling-based
method that employs the uniform distribution between a lower and an
upper bound to arrive at a randomly calculated chunk size between
these bounds)~\cite{Ciorba:2018} have been implemented in the LaPeSD
libGOMP\,\cite{LaPeSD-LibGOMP} based on the GNU \openmp library. 
The LLVM \openmp runtime\,\cite{LLVMOpenMP} also provides an implementation of TSS\,\cite{TSS:1993}
and \emph{static stealing} (also referred to as fixed-size chunking~\cite{FSC:1985}). 
FAC2 has also been recently implemented in
the LLVM \openmp runtime to offer further performance enhancement
possibilities at higher loads~\cite{Kasielke:2019}. 


This review of existing related efforts shows that there is a large
amount of ad-hoc development of loop scheduling strategies and
schedulers for \openmp in various \openmp runtime libraries (RTLs),
yet none of these efforts comply with the \openmp specification.
While these implementations may remain helpful to certain users,
applications, and systems, their broad practical usability may be limited,
rendering them not useful for supporting the development of novel
advanced loop scheduling strategies in \openmp.

The main challenge is to \emph{decouple} the loop
scheduling strategy from its implementation strategy.
Such a decoupling opens the door to a broad range of dynamic adaptive
loop scheduling strategies that simply cannot be efficiently
implemented in \openmp RTLs, such as adaptive weighted factoring~\cite{AWF:2003} and adaptive
factoring~\cite{AF:2000} that adapt to changes during execution; 
strategies that mix static and dynamic scheduling to maintain a balance between data locality and load balance\,\cite{dynwork,dynwork2}; 
and fault-tolerant and energy-oriented loop scheduling strategies that require domain-specific knowledge\,\cite{openmpft,openmplowpower}. 

%% file: rq-analysis.tex

Let us consider what is needed to specify an arbitrary scheduling strategy for
a parallel loop. The strategy can use a combination of shared data
structures, a collection of low-overhead steal work queues, exclusive
queues meant for each core, or shared queues from which multiple
threads can dequeue tasks each representing a chunk of loop iterations
of a parallel loop. To enable the ability to learn from recent execution
history, e.g., recent outer iterations, or to make decisions about the
scheduling strategy based on information from libraries handling
inter-node parallelism, e.g., slack from MPI
communication\,\cite{adagio}, the scheduling strategy needs the
ability to pass a call-site specific history-tracking
object\,\cite{slackSched}. 


To adapt a loop scheduling strategy's parameters, e.g.,
chunk size, we provide a mechanism for a UDS to store the history of loop
timings or other statistics across loop invocations in an
application program, e.g., a simulation time-step of a numerical
simulation. Such a mechanism improves productivity for the application
programmer. The adjustment of the loop
scheduling strategy during execution reduces the need for manual performance tuning and
compiler-guided performance tuning, which for certain applications
such as those involving sparse matrix vector multiplication is
difficult, and for other applications such as a galaxy simulation
involving an $N$-body computation, is nearly impossible.

\begin{figure}[t!]
	\centering
	\begin{tikzpicture}[scale=0.6, every node/.style={scale=1,font=\scriptsize}]
	\coordinate (origin) at (0,0);
	\tikzstyle{line} = [draw,thick]
	\tikzstyle{bstyle} = [rectangle, draw, fill=brown!60, text centered, rounded corners, minimum height=2.6em,text width={1.4cm}]
	\node [bstyle] at (1,0) (Init) {Setup};
	\node [bstyle,right=1cm of Init] (OnQueue) {Enqueue};
	\node [bstyle,right=1cm of OnQueue] (while) {While(todo ! empty)};
	\node [bstyle,right=1cm of while] (CleanUp) {Clean up};
	\node [bstyle,below =1cm of CleanUp] (DeQueue) {Enqueue};
	\node [bstyle,left=1cm of DeQueue] (BeginLoopBody) {Begin body};
	\node [bstyle,left=1cm of BeginLoopBody] (LoopBody) {Chunk of iterations};
	\node [bstyle,left=1cm of LoopBody] (EndLoopBody) {End body};
		
	\draw [-latex,line] ($(Init.east)+(0.0,0.0)$) -> ($(OnQueue.west)-(0.0,0.0)$);	
	\draw [-latex,line] ($(OnQueue.east)+(0.0,0.0)$) -> ($(while.west)-(0.0,0.0)$);	
	\draw [-latex,line] ($(while.south east)+(0.0,0.0)$) -- ($(BeginLoopBody.north east)+(0.0,1.0)$) -- ($(DeQueue.north)+(0.0,1.0)$) -> ($(DeQueue.north)-(0.0,0.0)$);	
	\draw [-latex,line] ($(DeQueue.west)+(0.0,0.0)$) -> ($(BeginLoopBody.east)-(0.0,0.0)$);	
	\draw [-latex,line] ($(BeginLoopBody.west)+(0.0,0.0)$) -> ($(LoopBody.east)-(0.0,0.0)$);	
	\draw [-latex,line] ($(LoopBody.west)+(0.0,0.0)$) -> ($(EndLoopBody.east)-(0.0,0.0)$);	
	\draw [-latex,line] ($(EndLoopBody.north)+(0.0,0.0)$) -- ($(EndLoopBody.north)+(0.0,0.5)$) -- ($(BeginLoopBody.north west)+(0.0,0.5)$) -> ($(while.south west)-(0.0,0.0)$);	
	\draw [-latex,line] ($(while.east)+(0.0,0.0)$) -> ($(CleanUp.west)-(0.0,0.0)$);
	\end{tikzpicture}
	\caption{Basic loop scheduler code structure.}
        \label{fig:loopschedstructure}\vspace*{-0.5cm}
\end{figure}
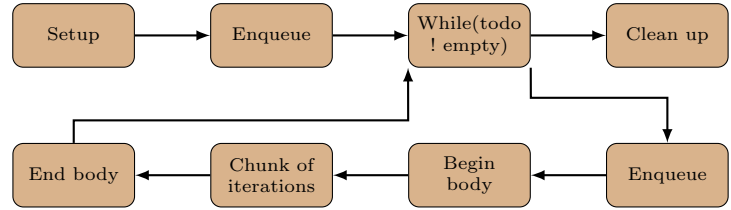

In order to support UDS in OpenMP, we must first understand
the principal components of loop scheduling. Figure~\ref{fig:loopschedstructure} shows a control flow diagram
of the basic loop scheduling code structure.
In principle, an OpenMP
loop scheduling problem can be represented as a \emph{todo list} of loop
iterations (or chunks of loop iterations), that must somehow be mapped
to parallel execution units. To manage such a \emph{todo list}, and
assuming an undefined initial state, three specific operations are
required:
\begin{enumerate}[(a)]
   \item \emph{a setup operation} to generate a known initial state,
   i.e., the todo list must be created and initialized,
   \item \emph{an enqueue operation} to place the loop iterations on the
   todo list, and 
   \item \emph{a dequeue operation} to select the next loop iteration to be
   executed from the todo list.
\end{enumerate}

As OpenMP requires that the precise iteration space is known before
the loop execution starts, the todo list is conceptually completely filled
at the beginning of loop execution with all the chunks of loop
iterations, and subsequently consumed by iterative dequeue operations
by each \openmp thread. The dequeue operation then implements \emph{an arbitrary
scheduling strategy} or pattern. Constraints, such as
sequential ordering or the scheduling pattern are solely an aspect of
the dequeue operation as well as any synchronization mechanisms to
maintain parallel safety of the used data structures. 
For both the enqueue and dequeue functions, the master thread can 
potentially serve a different function than the
remaining threads in a loop scheduler. Also, the behavior of
the threads needs to be specified either via
function pointers or declaratively. Such
specification must be done while preserving
generality so that novel loop scheduling strategies
have the ability to deal with the loop's
iteration space in a controlled manner.
As an example, we have shown how dynamic scheduling
can be optimized by using a combination of statically
scheduled and dynamically scheduled loop iterations\,\cite{dynwork2},
where the dynamic iterations still execute in consecutive order on
a thread to the extent possible\,\cite{dynwork6}. 

Good practice also recommends to clean up after performing work, as
the OpenMP base languages do not offer automatic garbage
collection. Hence, a clean-up, or post scheduling operation is needed.

Analyzing the current state of the art in loop scheduling in Section\,\ref{sec:bkgrnd-sota}, we identified three categories of strategies:
\begin{enumerate}[(1)]
\item \emph{static loop scheduling}: each thread is assigned a fixed workload,
\item \emph{dynamic non-adaptive loop scheduling}: each thread requests iterations according to a fixed pattern, and 
\item \emph{dynamic adaptive loop scheduling}: each thread requests iterations according to a variable pattern, while the performance of work chunks is measured and scheduling pattern is adjusted accordingly.
\end{enumerate}

For loop scheduling strategies of type (1) and type (2), in principle, only the three
operations are required. For strategies of type (3), the execution behavior
of previous iterations of the loop body is used as input to determine the
scheduling strategy parameters, e.g., next chunk size, 
to use for scheduling chunks of loop iterations of the current loop iteration and/or
invocation. To accommodate such scheduling strategies, a mechanism needs to be provided
to \textit{obtain} information during previous loop iterations and/or invocations and a
mechanism to \textit{store} this information.To obtain
the information, measurement facilities for the loop body may be required, be
it explicit operations, such as `begin-loop-body'--`end-loop-body' to allow
for measurements, or implicit facilities, e.g., as defined by the \openmp tools interface.

To store information, i.e., a form of execution history that must be preserved across dequeue operations
to account for past behavior, UDS must provide a mechanism to
store and access the history of loop timings or other statistics across
multiple loop iterations and/or invocations in an application program, e.g., across simulation time-steps 
of a numerical simulation.

With these functions and mechanisms, a user of OpenMP can declare in the code a
\texttt{schedule} clause of kind \texttt{X}. In the declaration, the user would specify a function
to \emph{initialize} the scheduler, a function to \emph{enqueue} chunks onto a
shared queue, a function to \emph{dequeue} chunks of iterations from a queue
by a thread, a function for garbage collection (\emph{finalize}) after loop scheduling is done,
and, optionally, \emph{begin} and \emph{end} functions for a dynamic adaptive loop scheduling. 
Then, function \texttt{X\_init()} allows a user-defined scheduling to
allocate and initialize its data structures that are to be used
commonly across parallel loops that use \texttt{X}. The functions \texttt{X\_enqueue()} and 
\texttt{X\_dequeue()} determine a loop's indices that a thread should work on based on the parameter
values for the scheduling strategy and the loop. Every thread in
the team should call \texttt{X\_dequeue()} repeatedly. For adaptive loop
scheduling, one needs to have an \texttt{X\_begin()} and \texttt{X\_end()} function 
for measurements of the current
invocation of a loop used for history used for adapting the parameters of the
scheduling strategy used in subsequent iterations and/or invocations of the loop. Finally, a user can optionally
define a data structure to store timings of a loop or other data to enable
persistence over invocations of an OpenMP parallel loop. 

As long as one is allowed to define the four functions (init, enqueue, dequeue, and finalize), together with the
begin and end functions for gathering per-loop invocation data and
data structure for storing history of the data, one can implement \emph{any
user-defined loop scheduling} through a loop scheduler. Formally, the four
functions together with begin and end functions and class
declaration and definition for the history object are \emph{necessary and
sufficient} to fully express an arbitrary user-defined loop scheduling strategy.

%% file: uds-proposal.tex
As described in Section\,\ref{sec:rq-analysis}, only six operations,
i.e., init, enqueue, dequeue, finalize, begin-loop-body, and end-loop-body
must be defined in order to implement all existing loop scheduling
strategies. While not all of those operations must be implemented by a
given loop scheduling strategy, it must be possible to implement those
operations. An interface for a UDS in \openmp must enable
such definitions from the user program without having to alter
the \openmp runtime library. However, due to a programmer's desire for
brevity, such an interface should avoid verbosity and enable
efficient and quick specification of new scheduling strategies.

Due to the restriction and requirements of the OpenMP language on
loops, the set of six operations can further be reduced. As the
iteration space of loops with OpenMP \texttt{parallel for} must be
fixed prior to loop execution, the enqueue function must only be
executed prior of the actual loop execution.
It, therefore, can be merged with the init operation.
The dequeue operation and the begin-loop-body operations are
executed, if defined, always back-to-back.
Hence, these operations can also be implemented in a single merged
operation. The conceptional code transformation
(see\,Fig.~\ref{fig:loopschedstructure}) in combination with a loop
similarly provides a way to merge the end-loop-body operation with the
dequeue operation.

This results in \emph{only three} operations that must be defined by a UDS
developer in the context of \openmp loop scheduling: a \emph{start} routine
implementing the setup and enqueue operation, a \emph{get-chunk} operation
implementing the end-body, dequeue and begin-loop-body operation,
and a \emph{finish call} for the finalize operation.

The concept of a todo list of loop iterations is rather impractical
for OpenMP loops, as the iteration space may be large and an explicit
enumeration of all iterations is not practical. 
Thus, the todo list is typically implemented as a set of shared or
thread-private loop counters.

For current implementations of \openmp parallelized loops in Intel,
LLVM and GNU Runtime Libraries, we observe a common implementation pattern.
Using the three fundamental operations of init, dequeue and finalize,
these compilers transform an \openmp `parallel for` as follows using the
following pattern: a setup operation, a while loop with a dequeue
function and a tailing end operation, which implements cleanup of
residual stack data (see code at the top of this page).
The three OpenMP loop scheduling strategies, i.e. static, guided, and
dynamic, are implemented using similar patterns\,\cite{Kasielke:2019}.
\begin{table}[t!]
\vspace{-.2cm}
\centering
\begin{tabular}{c c c}%
\begin{minipage}[t][4cm][c]{0.38\textwidth}%
\begin{lstlisting} 
#pragma omp parallel for
  for (i=0;i<iMax;i++)
	{
    ... LOOP BODY ...
  }
\end{lstlisting}
\end{minipage}%
&\begin{minipage}[t][4cm][c]{0.036\textwidth}$\rightarrow$\end{minipage}&
\begin{minipage}[t]{0.56\textwidth}
\begin{lstlisting}
#pragma omp parallel
{
  init(...);
  #pragma omp barrier
  while(!done){
    for (each item in dequeue(...))
      ... LOOP BODY ...
  }
  finalize(...);
}
\end{lstlisting}
\end{minipage}\\
\end{tabular}\vspace{-.5cm}\end{table}%
\vspace{-0.1cm}
A UDS specification must allow a loop scheduling implementer to access
critical loop parameters and program data: 
\begin{inparaenum}[\itshape a)\upshape]
\item lower bound,
\item upper bound,
\item stride,
\item custom data, e.g. loop history data or NUMA information, and
\item chunk size.
\end{inparaenum}
The `chunk size' here is not the \texttt{chunksize} parameter
frequently referred to in the \openmp{} \texttt{schedule()} clause,
but an optimization parameter used to group multiple iterations into a
single loop scheduling item. 

We currently propose two complementary proposals for an interface for
a UDS, enabling a user specification for those three functions. However, the
design of these interfaces substantially differs at the \openmp host
language level: 
\begin{inparaenum}[(1)]
\item a C++-geared interface using a concept similar to lambdas and 
\item a more classic C/Fortran-geared interface similar to user-defined reductions in \openmp.
\end{inparaenum} 
\vspace{-.25cm}
\subsection{Lambda-style Specification for UDS}
\label{sec:LambdaLike}
Using a lambda-style syntax, a scheduling implementer can define code
to implement the setup, dequeue, and finalize operations.
\vspace{-0.1cm}
\begin{lstlisting}
#pragma omp parallel for \ 
  schedule(UDS[:chunkSize, [monotonic| non-monotonic]) \
  [init(@@INIT_LAMDA@@)] dequeue(@@DEQUEUE_LAMDA@@)    \
  [finalize(@@FINISH_LAMDA@@)] [uds_data(void*)]
 \end{lstlisting}
\vspace{-0.1cm}
To access the critical loop parameters, we propose compiler-generated
getter and setter functions.
\vspace{-0.1cm}
\begin{lstlisting}
inline unsigned int OMP_UDS_loop_start();
inline unsigned int OMP_UDS_loop_end();
inline unsigned int OMP_UDS_loop_step();
inline unsigned int OMP_UDS_chunksize();
inline unsigned int OMP_UDS_user_ptr();
\end{lstlisting}
\vspace{-.15cm}
\begin{lstlisting}
void OMP_UDS_loop_chunk_start(int start_iteration); 
void OMP_UDS_loop_chunk_end(int end_iteration);
void OMP_UDS_loop_chunk_step(int step_size);
void OMP_UDS_loop_dequeue_done();
\end{lstlisting}
\vspace{-0.1cm}
To compile a loop scheduled using a UDS, the compiler mixes the lambda
code into the respective regions in the loop transformation pattern.
The setter and getter functions can furthermore be inlined and their
values propagated by constant value propagation, to further reduce and
optimize the specific loop code.

As this interface would require a definition for every use of a specific loop scheduling approach, a template-like directive defines
reusable schedules without the need to repeat the actual UDS code at every usage.

\vspace{-0.1cm}
\begin{lstlisting}
#pragma omp declare schedule_template (mystatic)  \
  [init(@@INIT_LAMDA@@)] dequeue(@@DEQUEUE_LAMDA@@) \
  [finalize(@@FINISH_LAMDA@@)] [uds_data(void*)]   
\end{lstlisting}
\vspace{-0.1cm}

\begin{lstlisting}
#pragma omp parallel for schedule(UDS,template(mystatic))
  for (int i = 0; i < n; i++) 
  { ... LOOP BODY ... }
  \end{lstlisting}
\vspace{-0.1cm}

The availability of both UDS templates and localized UDS allows for
implementation of libraries supported UDSs, but preserves the ability to either
specify localized single use loop scheduling strategies or to
overwrite specific elements of an existing UDS template for a specific
loop.
An example of how the user could implement the above \verb|mystatic| is provided in Fig.~\ref{fig:sd} where the left side illustrates a naive implementation of the \openmp \verb|static| scheduling clause using  \textit{lambda-style UDS} based on the \texttt{chunksize} specified by the programmer.
\vspace{-.25cm}
\subsection{Specifying UDS via \texttt{declare} Directives}
\label{sec:UDRSP}
The second variant for specifying UDS derives from the existing syntax
for a user-defined reduction, or UDR, in \openmp.
Here, the \texttt{declare schedule} clause defines a new named
scheduling using user-defined functions with positional arguments:
\vspace{-.1cm}
\begin{lstlisting}
#pragma omp declare schedule(mystatic) arguments(2) \
  init(my_init(omp_lb, omp_ub, omp_inc, omp_arg0, omp_arg1)) \
  next(my_next(omp_lb_chunk, omp_ub_chunk, omp_arg0, omp_arg1)) \
  fini(my_fini(omp_arg1))
  \end{lstlisting}

\vspace{-.05cm}
The \verb|arguments| sub-clause allows to specify the number of additional arguments beyond
the required arguments. The reserved
keywords \verb|omp_lb|, \verb|omp_ub|, \verb|omp_inc|, \verb|omp_lb_chunk|, 
and \verb|omp_ub_chunk| serve as markers for the compiler what
information about the loop iteration space to pass to the UDS, as the
user code expects this information as a function argument.
The compiler generates \verb|omp_arg0| .. \verb|omp_argN| as
necessary, based on the count in the \verb|arguments| \mbox{sub-clause}.
However, the \openmp-defined arguments must always be the first
arguments, followed by any user-defined arguments.
This allows, for example, simpler scheduling strategies to omit unused
information.
The additional user-provided arguments use the type of the argument at
the use-site of the \mbox{user-defined} scheduling, similar to
the \verb|auto|-type in C++. The function implementations must then
use the appropriate types and provide the implementation of the scheduling strategy. 
Please note, that a definition of the \verb|next| function must return a non-zero value
if unprocessed loop chunks remain, and zero if the loop has been 
completed.
\vspace{-.1cm}
\begin{lstlisting}
void mystatic_init(int lb, int ub, int inc, loop_record_t * lr);
int mystatic_next(int * lower, int * upper, loop_record_t * lr);
void mystatic_fini(loop_record_t * lr);
\end{lstlisting}
\vspace{-.1cm}


To generate code from such a UDS specification, the compiler employs the
standard loop transformation pattern it uses today and replaces the
calls to its scheduling function with user-supplied functions of the
UDS. The compiler may then match the types defined by the scheduling
implementing function definitions to generate error messages, if a
type mismatch is detected, or apply inlining to remove the function
call.

The following example showcases how a user-defined scheduling
strategy would be used and how parameters are passed to the scheduler:
\begin{lstlisting}
#pragma omp parallel for schedule(mystatic(&lr))
  for (i = 0; i < sz; i++) {
#pragma omp atomic
    array[i]++;
  }
}
\end{lstlisting}

An example of how the user could implement the above schedule \verb|mystatic| is provided in Fig.~\ref{fig:sd}, where the right side shows a naive implementation of the \openmp \verb|static| scheduling clause using \textit{declare-style UDS} based on the \texttt{chunksize} specified by the programmer.
\begin{figure}[ht!]
\vspace{-.2cm}
\centering

\begin{tabular}{c c c}%
\begin{minipage}[t][\textheight-4cm]{0.495\textwidth}%
\begin{lstlisting}[basicstyle=\scriptsize\ttfamily]
typedef struct {
  int * next_lb;
} loop_record_t;

void mystatic_init() {
 int tid = omp_get_thread_num();
 #pragma omp single 
 {
  OMP_UDS_user_ptr()->next_lb = 
   malloc(sizeof(int)*omp_get_num_threads());	
 }
 OMP_UDS_user_ptr()->next_lb[tid] =
  lb+tid * chunksz;
}

void mystatic_next() {
 int tid = omp_get_thread_num();
 if (OMP_UDS_user_ptr()->next_lb[tid] >= 
  OMP_UDS_loop_end()) return 0;
 OMP_UDS_loop_chunk_start(
  OMP_UDS_user_ptr()->next_lb[tid]);
 if (OMP_UDS_user_ptr()->next_lb[tid] +
  OMP_UDS_chunksize() >= 
  OMP_UDS_loop_end()) {
  OMP_UDS_loop_chunk_end(OMP_UDS_loop_end());		
  } 
  else {
   OMP_UDS_loop_chunk_end(
    OMP_UDS_user_ptr()->next_lb[tid] + 
    OMP_UDS_chunksize());
  }
 OMP_UDS_user_ptr()->next_lb[tid] =
  OMP_UDS_user_ptr()->next_lb[tid] +
  omp_get_num_threads()*OMP_UDS_chunksize();
 OMP_UDS_loop_chunk_step(
  OMP_UDS_loop_step());
 return 1;
}

void mystatic_fini(){
 free(OMP_UDS_user_ptr()->next_lb);
}
					
#pragma omp declare \
 schedule_template(mystatic)\
 init(mystatic_init())\
 next(mystatic_next())\
 finalize(mystatic_fini()) 
\end{lstlisting}
\end{minipage}%
&&
\begin{minipage}[t][\textheight-4cm]{0.495\textwidth}

\begin{lstlisting}[basicstyle=\scriptsize\ttfamily]
typedef struct {
 int lb;
 int ub;
 int incr;
 int chunksz;
 int * next_lb;
} loop_record_t;

void mystatic_init(int lb, int ub, int incr,
     int chunksz,loop_record_t * lr) {
 int tid = omp_get_thread_num();
 #pragma omp single 
 {
  lr->lb = lb;
  lr->ub = ub;
  lr->incr = incr;
  lr->next_lb = malloc(sizeof(int)*
	  omp_get_num_threads());
  lr->chunksz = chunksz;
 }
 lr->next_lb[tid] = lb + tid * chunksz;
}

int mystatic_next(int * lower, int * upper,
    int * incr, loop_record_t * lr) {
 int tid = omp_get_thread_num();
 if (lr->next_lb[tid] >= lr->ub) return 0;
 *lower = lr->next_lb[tid];
 if (lr->next_lb[tid] +
   lr->chunksz >= lr->ub)
  *upper = lr->ub;		
 else
  *upper = lr->next_lb[tid] + lr->chunksz;
 lr->next_lb[tid] = lr->next_lb[tid] +
   omp_get_num_threads()*lr->chunksz;
 *incr = lr->incr;
 return 1;
}

int mystatic_fini(loop_record_t * lr) {
 free(lr->next_lb);
}	
		
#pragma omp declare schedule(mystatic) \
 arguments(1) init(mystatic_init(omp_lb, \
 omp_ub,omp_incr,omp_chunksz,omp_arg0) \
 next(mystatic_next(omp_lb_chunk, \
 omp_ub_chunk,omp_chunk_incr,imp_arg0)) \
 fini(mystatic_fini(imp_arg0)
\end{lstlisting}
\end{minipage}\\
\end{tabular}\vspace{-1.0cm}
	\caption{Naive example for implementing the \openmp \texttt{static} scheduling clause using both proposed UDS strategies. Left side presents the implementation following the \textit{lambda-style} specification, Sec.~\ref{sec:LambdaLike}, while the right side follows the \textit{declare-directives} style, Sec.~\ref{sec:UDRSP}.}
	\label{fig:sd}
\vspace{-.5cm}
\end{figure}%

%% file: discussion.tex
We consider both proposals sufficient as a UDS specification layer. As \openmp targets three separate host languages, we must consider the implications of each interface to the host language and use in daily programming work.

The \textit{lambda-style} interface easily fits into the language canon of C++, where the concept of lambdas already exists and can easily be reused in the context of UDS. 
Also, the use of getter and setter functions does not present a source of overhead, as existing compiler optimizations, such as inlining and constant-value propagation and folding, will enable removal of all explicit function calls. 
As some operations, i.e., setup and finalize, are also not required for all implementations of a UDS, this avoids the verbose, potentially empty argument list of positional arguments, required by the second proposal. 
However, the flexibility and ease of iteration in C++ conflicts with C and Fortran, where lambda constructs are not (yet) available. 
While the concept of lambdas is likely to be added to Fortran in the future, the specific syntax and semantics are currently not known. 
At this point, we are also not aware of any efforts to add lambdas to C.
The \textit{UDR-style} specification has, in principle, a precedence-case in the UDR specification in \openmp.
While this approach relies on a more frumpy fixed position syntax style, it remains compatible with all three \openmp host languages.

A potential solution would allow the use of the lambda-style syntax for C++, and the UDR-style for C and Fortran codes.

%% file: relatedWork.tex
Our suggested UDS approach for supporting novel loop scheduling strategies
and two alternative interfaces for it have much
work related to it, which we mention here to distinguish our idea and
its development from the existing work.
Work on an \openmp runtime scheduling\,\cite{Automatic-OMP-LS:2012,RuntimeEmpSched} system
automatically chooses the schedule. The problem with this scheme is that it does not work for all application-architecture pairs: it allows no domain knowledge or architecture knowledge to be incorporated into it, which only a user would know. 
Methods such as setting the schedule of an OpenMP loop to `\texttt{auto}' are insufficient because the methods do not allow a user to take control of any decision of loop scheduling that the \openmp RTL makes\,\cite{dynwork7}. The emergence of threaded runtimes such as Argobots\,\cite{argobots} and QuickThreads\,\cite{quickThreads} are frameworks containing novel loop scheduling strategies, and they actually argue in favor of a flexible specification of scheduling strategies. In comparison, our work on the UDS specification is the first proposal that works at the \openmp standard specification level.

\comments{
Many frameworks for Runtime empirical  scheduling\,\cite{RuntimeEmpSched} \todoVK{@Vivek: Please write 
something about\,\cite{RuntimeEmpSched} since you may know it better than I do.} \\
   2. Autotuned loop scheduling by Vivek\,\cite{dynwork7}. \todoVK{@Vivek: Please add include and
   describe here your iWomp15 paper \,\cite{dynwork7}} \\
}

%% file: conclusion.tex
\openmp's loop scheduling choices do not always offer the best performance,
and standardization of all existing scheduling strategies is
infeasible. In this work, we showed that an OpenMP standard-compliant interface
is needed to implement an arbitrary user-defined loop scheduling
strategy. We presented two competing standard-compliant UDS interface proposals to support this
need. We conceptually compare the two proposed UDS interfaces in terms
of feasibility and capabilities regarding the programming languages C,
C++, and Fortran that host \openmp.

The immediate next step is the implementation of the UDS interfaces as a
prototype in an open source compiler, such as GNU or LLVM, to explore
the performance-related capabilities and benefits of the proposed approaches. 
As the Intel and LLVM OpenMP RTLs offer schedules choices beyond those in the \openmp standard,
we will work to expose those schedules using either or both UDS
proposals and evaluate their practical use for various application-architecture pairs. 
We welcome and value the feedback from the \openmp
community as we proceed in this direction.

%% file: ack.tex
\section*{Acknowledgments}
We thank Alice Koniges from Maui HPCC for providing us with NERSC's
cluster Cori for experimenting with machine learning applications
using \openmp, which helped us consider a relevant platform for user-defined
scheduling.
This work is partly funded by the Hessian State Ministry of Higher
Education by granting the ``Hessian Competence Center for High
Performance Computing" and by the Swiss National Science Foundation in
the context of the ``Multi-level Scheduling in Large Scale High
Performance Computers" (MLS) grant, number 169123.
\comments{
\footnotesize{
Intel and Xeon are trademarks or registered trademarks of Intel Corporation or its subsidiaries in the United States and other countries.

* Other names and brands are the property of their respective owners.

Software and workloads used in performance tests may have been
optimized for performance only on Intel microprocessors.  Performance
tests, such as SYSmark and MobileMark, are measured using specific
computer systems, components, software, operations and functions.  Any
change to any of those factors may cause the results to vary.  You
should consult other information and performance tests to assist you
in fully evaluating your contemplated purchases, including the
performance of that product when combined with other products.  For
more information go to \url{http://www.intel.com/performance}.

Intel's compilers may or may not optimize to the same degree for
non-Intel microprocessors for optimizations that are not unique to
Intel microprocessors. These optimizations include SSE2, SSE3, and
SSSE3 instruction sets and other optimizations. Intel does not
guarantee the availability, functionality, or effectiveness of any
optimization on microprocessors not manufactured by
Intel. Microprocessor-dependent optimizations in this product are
intended for use with Intel microprocessors. Certain optimizations not
specific to Intel microarchitecture are reserved for Intel
microprocessors. Please refer to the applicable product User and
Reference Guides for more information regarding the specific
instruction sets covered by this notice.
}
}